\documentclass[Afour,sageh,times]{sagej}

\usepackage{moreverb,url}
\usepackage[colorlinks,bookmarksopen,bookmarksnumbered,citecolor=blue,urlcolor=blue]{hyperref}
\newcommand\BibTeX{{\rmfamily B\kern-.05em \textsc{i\kern-.025em b}\kern-.08em
T\kern-.1667em\lower.7ex\hbox{E}\kern-.125emX}}
\def\volumeyear{2025}
\begin{document}
\title{Weibull Processes in Network Degree Distributions}
\runninghead{Williams and Chen}
\author{Peter R. Williams\affilnum{1,2} and Zhan Chen\affilnum{3}}
\affiliation{\affilnum{1}Rinna KK, Tokyo, Japan\\\affilnum{2}Independent Researcher\\\affilnum{3}Microsoft Japan, Tokyo, Japan}
\email{prw20042004@yahoo.co.uk}

\begin{abstract}
This study examines degree distributions in two large collaboration networks: the Microsoft Academic Graph (1800-2020) and Internet Movie Database (1900-2020), comprising $2.72 \times 10^8$ and $1.88 \times 10^6$ nodes respectively. Statistical comparison using $\chi^2$ measures showed that Weibull distributions fit the degree distributions better than power-law or log-normal models, especially at later stages in the network evolution. The Weibull shape parameters exhibit notable stability ($k \approx 0.8$-$1.0$ for academic, $k \approx 0.9$-$1.1$ for entertainment collaborations) despite orders of magnitude growth in network size. While early-stage networks display approximate power-law scaling, mature networks develop characteristic flattening in the low-degree region that Weibull distributions appear to capture better. In the academic network, the cutoff between the flattened region and power-law tail shows a gradual increase from $5$ to $9$ edges over time, while the entertainment network maintains a distinctive degree structure that may reflect storytelling and cast-size constraints. These patterns suggest the possibility that collaboration network evolution might be influenced more by constraint-based growth than by pure preferential attachment or multiplicative processes.
\end{abstract}

\maketitle

\section{Introduction}

The degree distribution $P(d)$ of a network characterises its connectivity structure, encoding essential information about network formation and evolution. In many real-world networks, these distributions have been found to follow power laws, where $P(d) \propto d^{-\gamma}$ \citep{barabasi1999}. This pattern appears across remarkably diverse systems, from scientific collaboration networks \citep{newman2001} to protein interactions \citep{jeong2001} and the World Wide Web \citep{albert1999}.

The canonical explanation for power-law degree distributions stems from preferential attachment mechanisms \citep{barabasi1999}, where the probability of a new node connecting to an existing node is proportional to the latter's degree. This ``rich-get-richer'' dynamic generates a scale-free network structure. However, \citet{broido2019} demonstrated that true scale-free networks are extremely rare, finding that only 4\% of studied networks showed strong evidence for power-law behaviour. \citet{voitalov2019} further showed that even when power-laws appear to fit well, they often fail rigorous statistical testing. These critiques have sparked renewed interest in alternative models for degree distributions.

Systematic deviations from power-law behaviour manifest in two key ways. First, in the low-degree regime, many networks exhibit characteristic flattening \citep{seshadri2021}, suggesting the presence of constraints on node connectivity. Second, in the high-degree tail, exponential cutoffs are common \citep{clauset2009}, implying natural limits to node growth. These deviations appear pronounced in social and biological networks, where physical or cognitive constraints may naturally limit connection formation.

Log-normal distributions, arising from multiplicative growth processes where $dk/dt \propto k$ \citep{mitzenmacher2004}, represent one alternative model. While these distributions can appear similar to power laws over limited ranges, they predict different growth mechanisms. Recent work has shown that both technological and biological networks may be better described by such processes \citep{wang2021}. However, log-normal models still struggle to capture the complex behaviour observed in many real networks, particularly in capturing both core and tail behaviour simultaneously.

Given the observed deviations from power-law behavior and the presence of apparent resource constraints, we consider the Weibull distribution as an alternative model that can capture both scale-free-like behavior and natural cutoffs. This distribution takes the form
\begin{equation}
P(d) = \frac{k}{\lambda}\left(\frac{d}{\lambda}\right)^{k-1}e^{-\left(d/\lambda\right)^k},
\label{eq:weibull}
\end{equation}
where $d$ is the degree, $k$ is a shape parameter and $\lambda$ is a characteristic scale parameter. Originally developed to model failure processes \citep{weibull1951}, this distribution has several appealing properties for network modelling. First, it naturally incorporates both power-law-like behaviour and exponential cutoffs through its shape parameter $k$. Second, it connects to competitive growth processes under resource constraints \citep{thompson2022}, aligning with our understanding of real-world network formation. Third, it has successfully described various complex systems, from firm-size distributions \citep{cabral2003} to income inequality \citep{ghosh2014} and species abundance patterns \citep{nekola2003}.

Understanding the precise functional form of degree distributions may be informative for several reasons. First, it provides insight into the fundamental mechanisms driving network formation and evolution. Second, different functional forms suggest different underlying constraints and growth processes, which may inform both theoretical models and practical applications. Third, accurate characterisation of degree distributions enables better prediction of network behaviour and more effective network design strategies. Fourth, identifying the correct distribution helps avoid the pitfalls of assuming universal mechanisms, as highlighted by recent critiques of the scale-free paradigm \citep{holme2019}.

The relationship between network structure and underlying mechanisms presents a fundamental challenge: while theoretical models can generate precise functional forms, real-world networks rarely conform to these idealized predictions. Rather than starting from theoretical models and testing their predictions, this study takes an empirical approach by first carefully measuring how degree distributions evolve in extensive longitudinal datasets. This observational strategy allows patterns to emerge from the data without presuming specific growth mechanisms. The subsequent fitting of different functional forms—derived from common theoretical models serves not to validate any single model, but rather to investigate the possible influence of different growth processes. 

This approach acknowledges that real networks likely emerge from multiple competing mechanisms operating simultaneously across different scales and time periods. By maintaining this empirically-grounded perspective, comparing fits across multiple candidate distributions, and examining how these fits evolve over time, we can identify which theoretical mechanisms appear most influential while remaining sensitive to the limitations of functional fitting. Such an approach is valuable for collaboration networks, where social, institutional, and practical constraints may create complex patterns that simple growth models cannot fully capture.

In this paper, we analyse the degree distribution evolution of two large collaboration networks: the Microsoft Academic Graph (1800-2020) and the Internet Movie Database (1900-2020). Our primary goal in this paper is to test empirically how well the three distributions mentioned above--— power-law, log-normal, and Weibull—-- capture the observed degree distributions in large, real-world collaboration networks over their centuries of evolution. We do not attempt to derive these distributions from a first-principles growth model here; rather, we focus on fitting these candidate functions to data and assessing goodness-of-fit.

\section{Methods}

\subsection{Dataset Overview}

The analysis examines two large collaboration networks: the Microsoft Academic Graph (MAG) \citep{magweb, sinha2015} and the Internet Movie Database (IMDb) \citep{imdbweb}. The MAG network comprises $2.72 \times 10^8$ authors connected by $1.8 \times 10^9$ edges representing co-authorship on $2.64 \times 10^8$ papers (1800-2020). The IMDb network contains $1.88 \times 10^6$ actors connected by $1.8 \times 10^6$ edges representing co-appearances in $6.34 \times 10^5$ movies (1900-2020). For each collaboration with $n$ contributors, edges were created between all possible pairs, generating $n(n-1)/2$ connections. The temporal evolution was tracked using a two-year collaboration window, with edges added two years before publication/release date and removed upon publication/release. This temporal framework was validated through sensitivity analyses using different collaboration durations and time-binning windows, with findings remaining robust across these variations. This analysis extends on the results presented in \citet{williams2025}.

\subsection{Degree Distribution Analysis}

For each network, degree distributions were computed yearly throughout their evolution. The degree distributions were constructed using an adaptive logarithmic binning scheme, with bin edges dynamically determined to maintain 1000 observations per bin. This approach, building on methods described in \citep{newman2005}, ensures consistent statistical power across the entire distribution while naturally adapting bin sizes to data density. In the low-degree region, where data points are abundant, this results in narrow bins that preserve fine-grained structure. In the high-degree tail, where data becomes sparse, the bins automatically widen to maintain statistical reliability. This dynamic binning method is particularly important for heavy-tailed distributions \citep{stumpf2012}, where traditional fixed-width bins can either obscure important features in dense regions or become unreliable in sparse regions. $\chi^{2}$ values were computed between the model fits and binned data, as a measure of goodness-of-fit.

Three distributions were fitted to the data: a power law $P(d) \propto d^{-\gamma}$ where $d$ is the degree, a log-normal distribution
\begin{equation}
P(d) = \frac{1}{d\sigma\sqrt{2\pi}} e^{-(\ln d - \mu)^2/2\sigma^2},
\end{equation}
and the Weibull distribution as described in Equation \ref{eq:weibull}. For power-law fits, maximum likelihood estimation following methods described by \citet{clauset2009} were employed. For log-normal and Weibull distributions, non-linear least-squares fitting with the Levenberg-Marquardt algorithm were used. The probability density functions were derived from the computed cumulative distributions through numerical differentiation.

\begin{figure*}[htb]
  \centering
  \begin{tabular}{cc}
    \includegraphics[width=0.47\textwidth]{"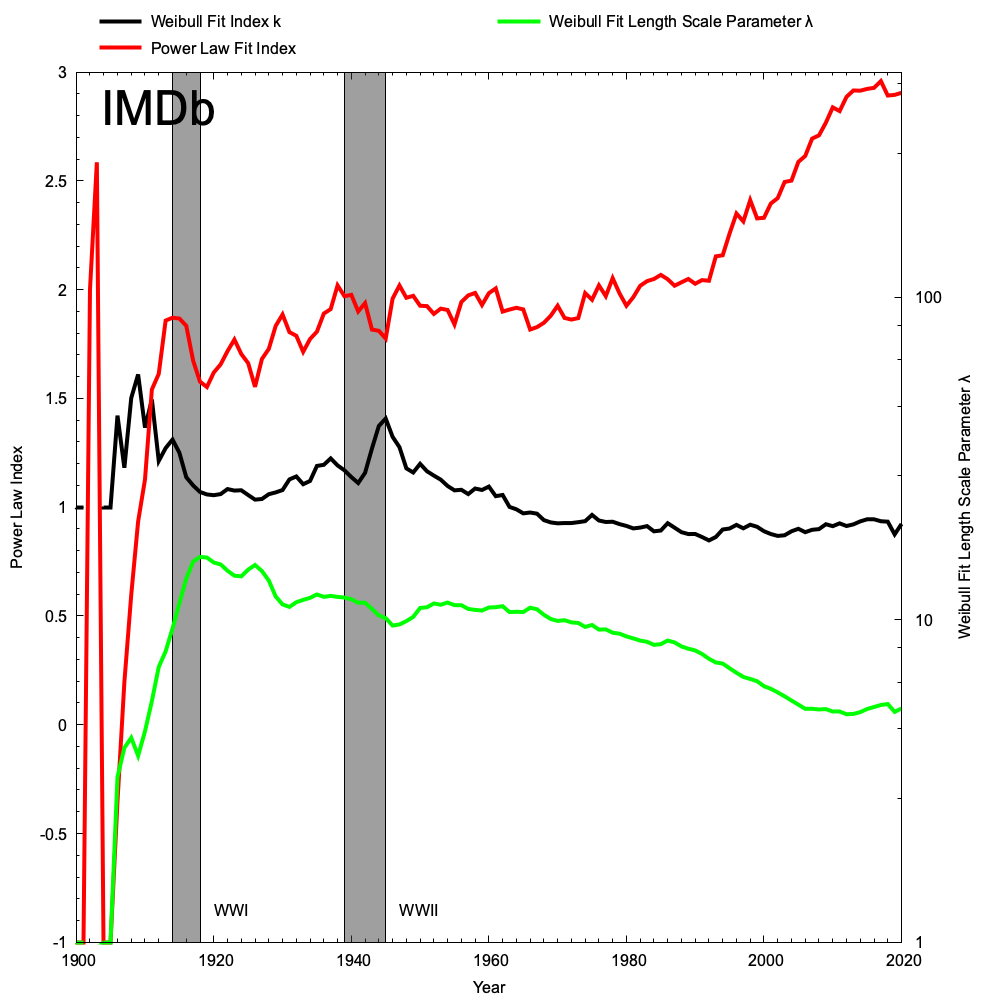"} &
    \includegraphics[width=0.47\textwidth]{"degree-distribution-parameters.png"}
  \end{tabular}
  \caption{Fitting parameters for power-law, log-normal, and Weibull fits to the MAG and IMDb network degree distributions in red, green, and black lines respectively.}
  \label{degree_fits}
\end{figure*}

\section{Results}

\subsection{Evolution of Degree Distributions}

\begin{figure*}[pt]
  \centering
  \begin{tabular}{c}
    \includegraphics[width=0.96\textwidth]{"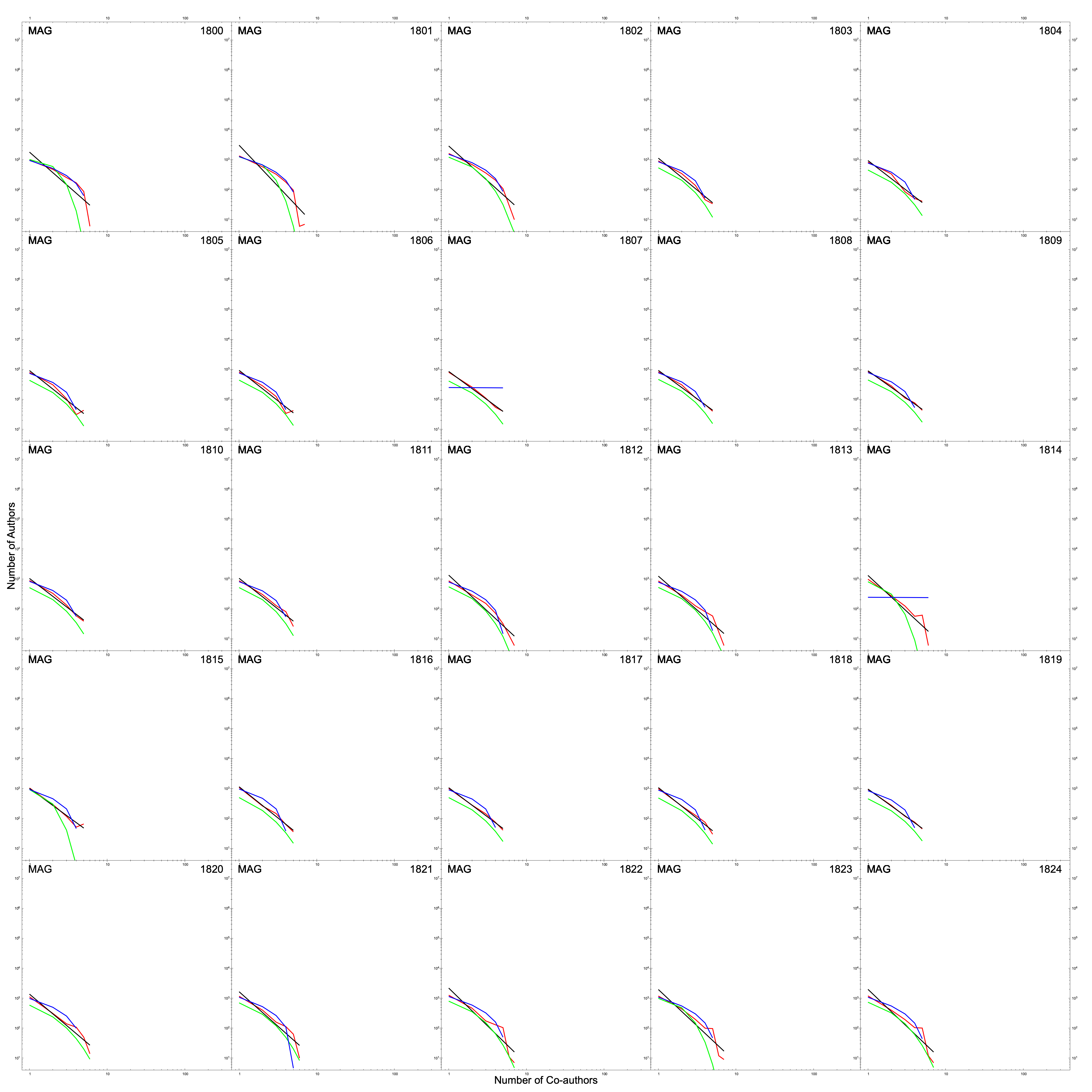"}
  \end{tabular}
  \caption{Degree distributions for cohorts of authors who first published in a given year, 1800 to 1824 (red line). Power-law, Log-normal, and Weibull fits are shown with black, blue, and green lines respectively.}
  \label{mag_degrees_a}
\end{figure*}

\begin{figure*}[pt]
  \centering
  \begin{tabular}{c}
    \includegraphics[width=0.96\textwidth]{"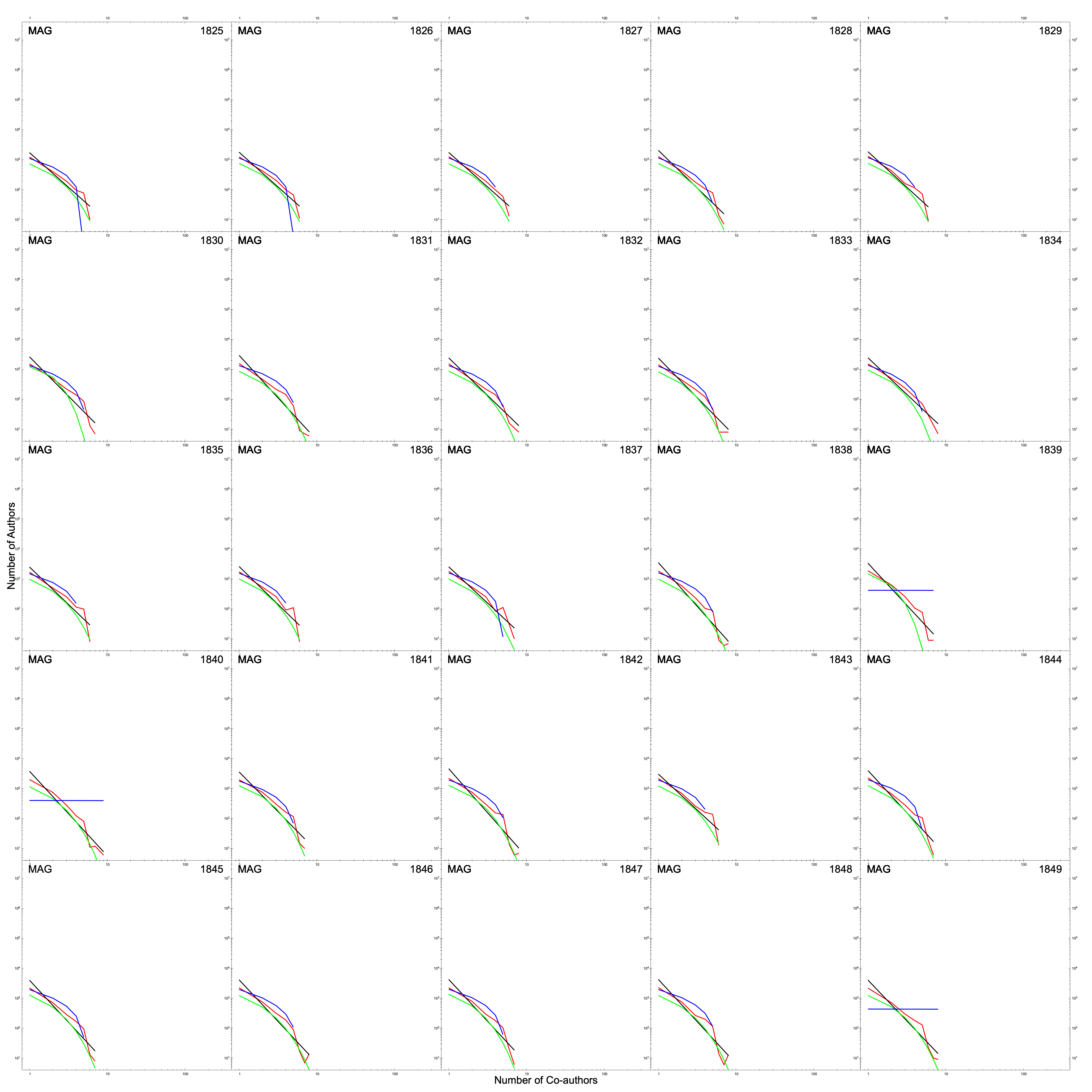"}
  \end{tabular}
  \caption{Degree distributions for cohorts of authors who first published in a given year, 1825 to 1849 (red line). Power-law, Log-normal, and Weibull fits are shown with black, blue, and green lines respectively.}
  \label{mag_degrees_b}
\end{figure*}

\begin{figure*}[pt]
  \centering
  \begin{tabular}{c}
    \includegraphics[width=0.96\textwidth]{"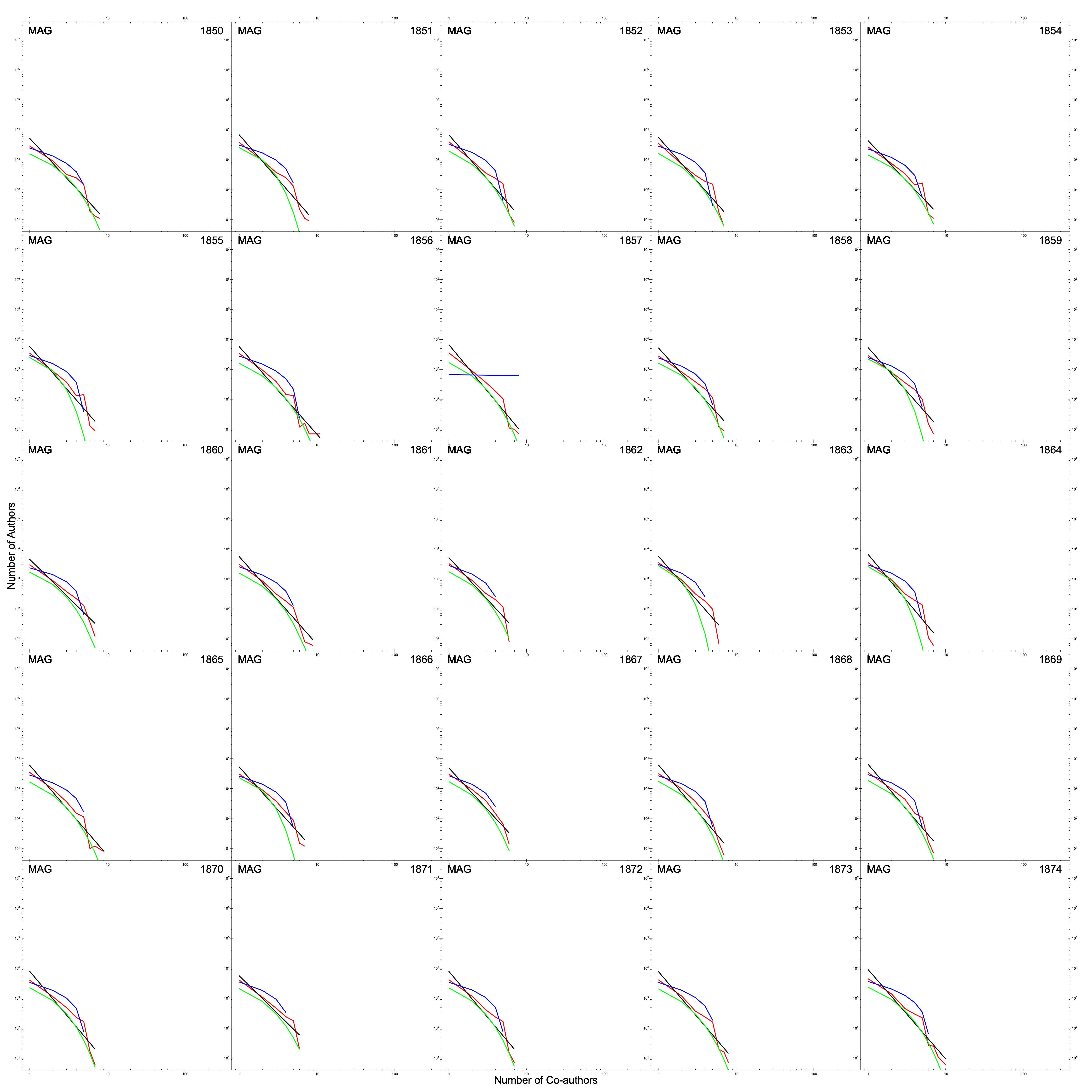"}
  \end{tabular}
  \caption{Degree distributions for cohorts of authors who first published in a given year, 1850 to 1874 (red line). Power-law, Log-normal, and Weibull fits are shown with black, blue, and green lines respectively.}  
  \label{mag_degrees_c}
\end{figure*}

\begin{figure*}[pt]
  \centering
  \begin{tabular}{c}
    \includegraphics[width=0.96\textwidth]{"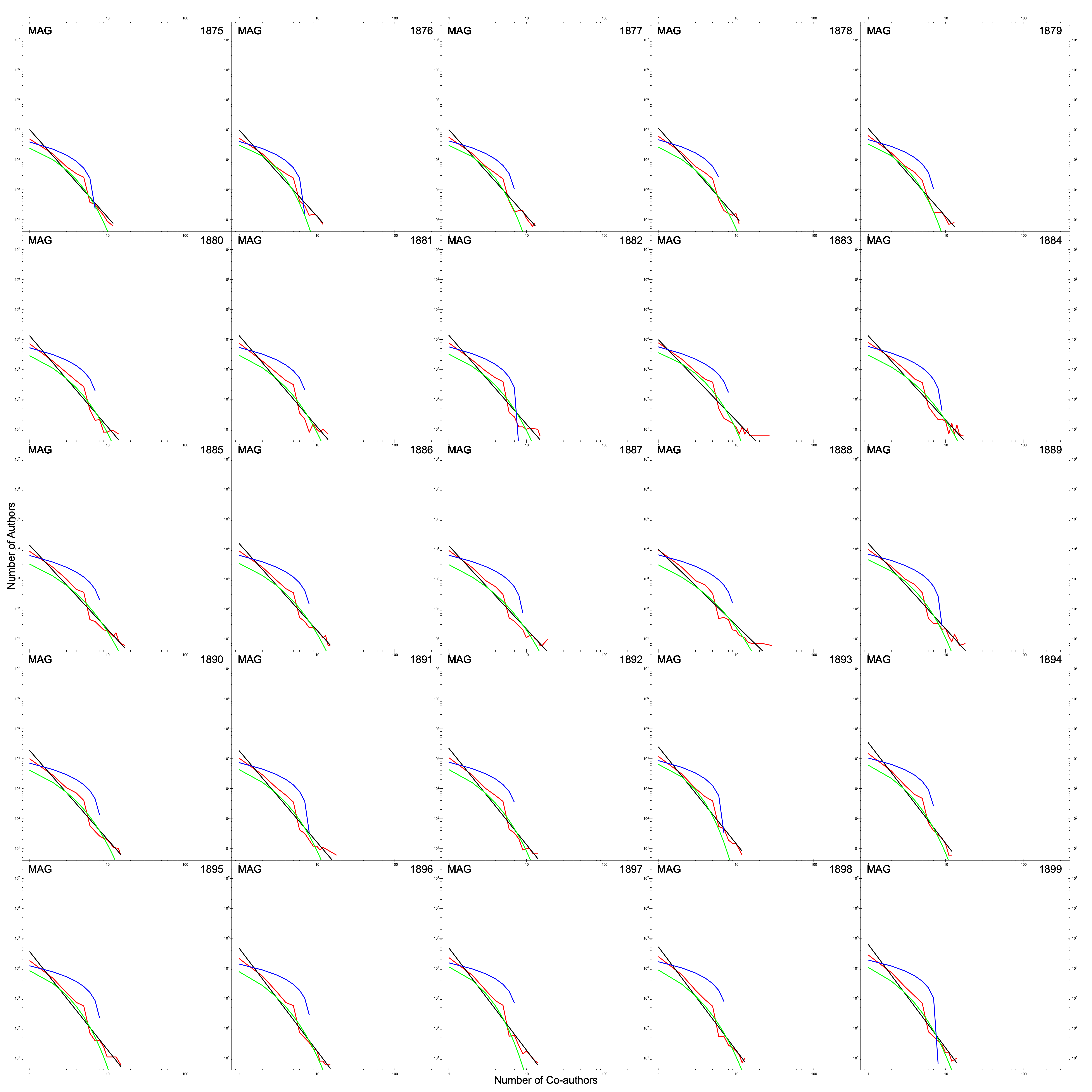"}
  \end{tabular}
  \caption{Degree distributions for cohorts of authors who first published in a given year, 1875 to 1899 (red line). Power-law, Log-normal, and Weibull fits are shown with black, blue, and green lines respectively.}
  \label{mag_degrees_d}
\end{figure*}

\begin{figure*}[pt]
  \centering
  \begin{tabular}{c}
    \includegraphics[width=0.96\textwidth]{"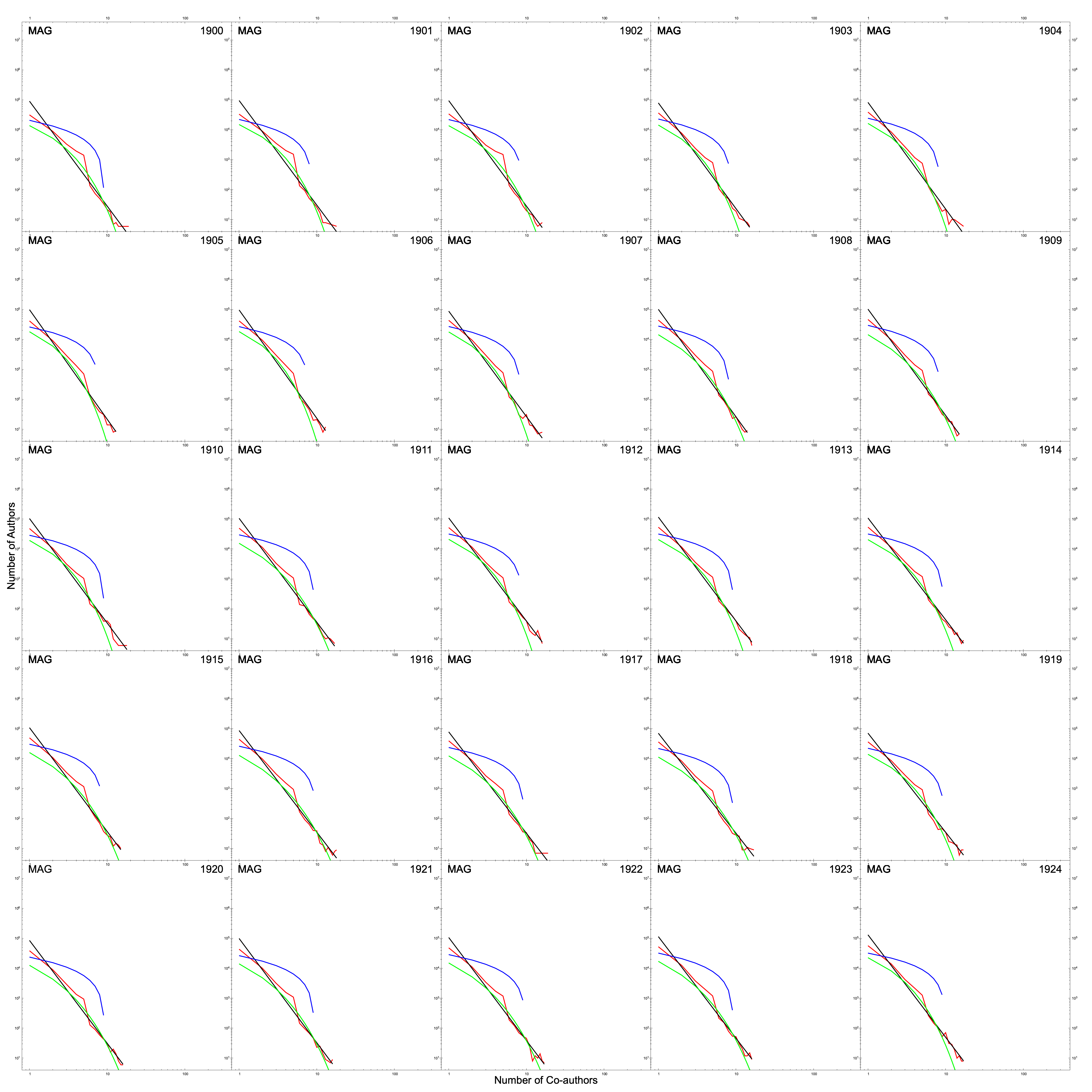"}
  \end{tabular}
  \caption{Degree distributions for cohorts of authors who first published in a given year, 1900 to 1924 (red line). Power-law, Log-normal, and Weibull fits are shown with black, blue, and green lines respectively.}
  \label{mag_degrees_e}
\end{figure*}

\begin{figure*}[pt]
  \centering
  \begin{tabular}{c}
    \includegraphics[width=0.96\textwidth]{"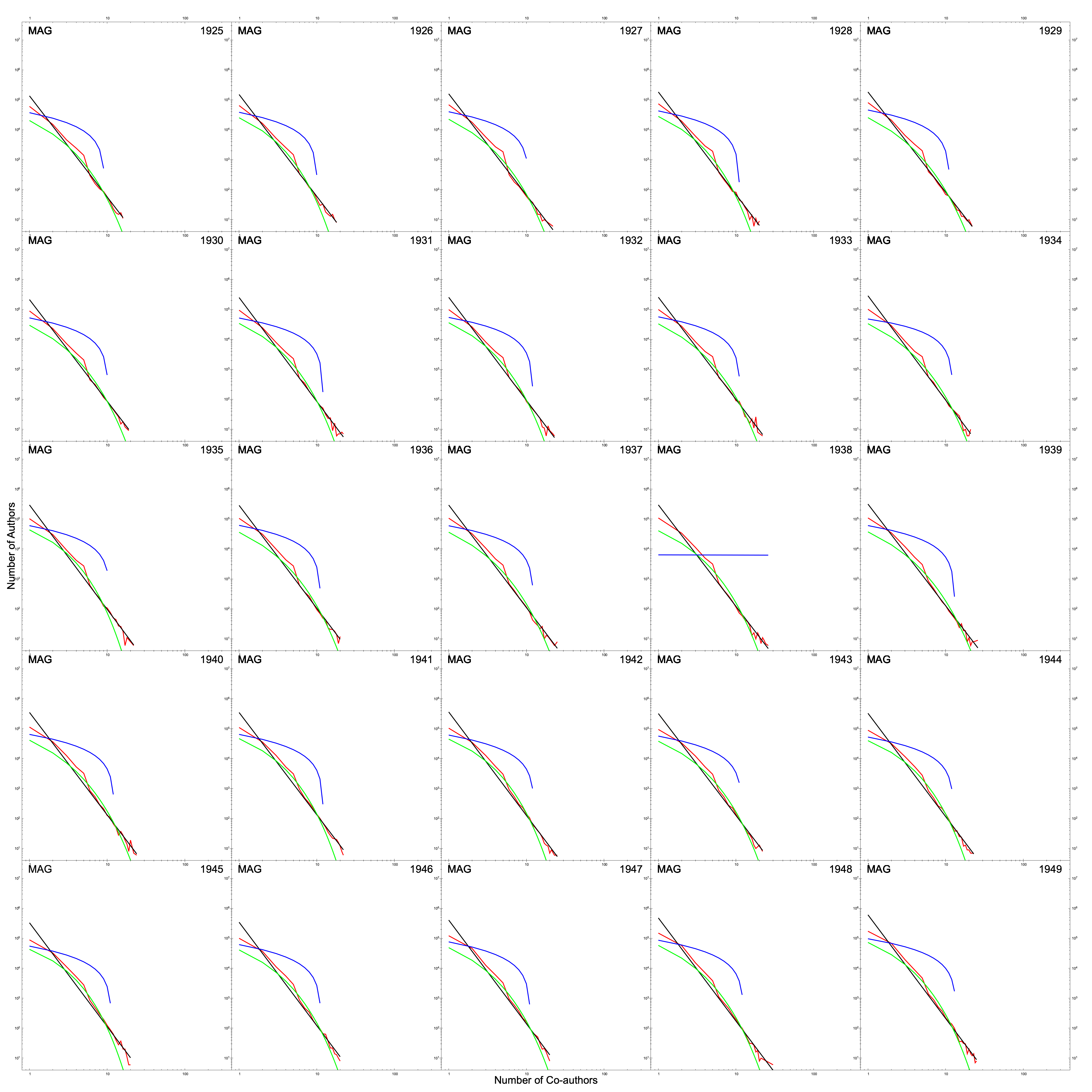"}
  \end{tabular}
  \caption{Degree distributions for cohorts of authors who first published in a given year, 1925 to 1949 (red line). Power-law, Log-normal, and Weibull fits are shown with black, blue, and green lines respectively.}
  \label{mag_degrees_f}
\end{figure*}

\begin{figure*}[pt]
  \centering
  \begin{tabular}{c}
    \includegraphics[width=0.96\textwidth]{"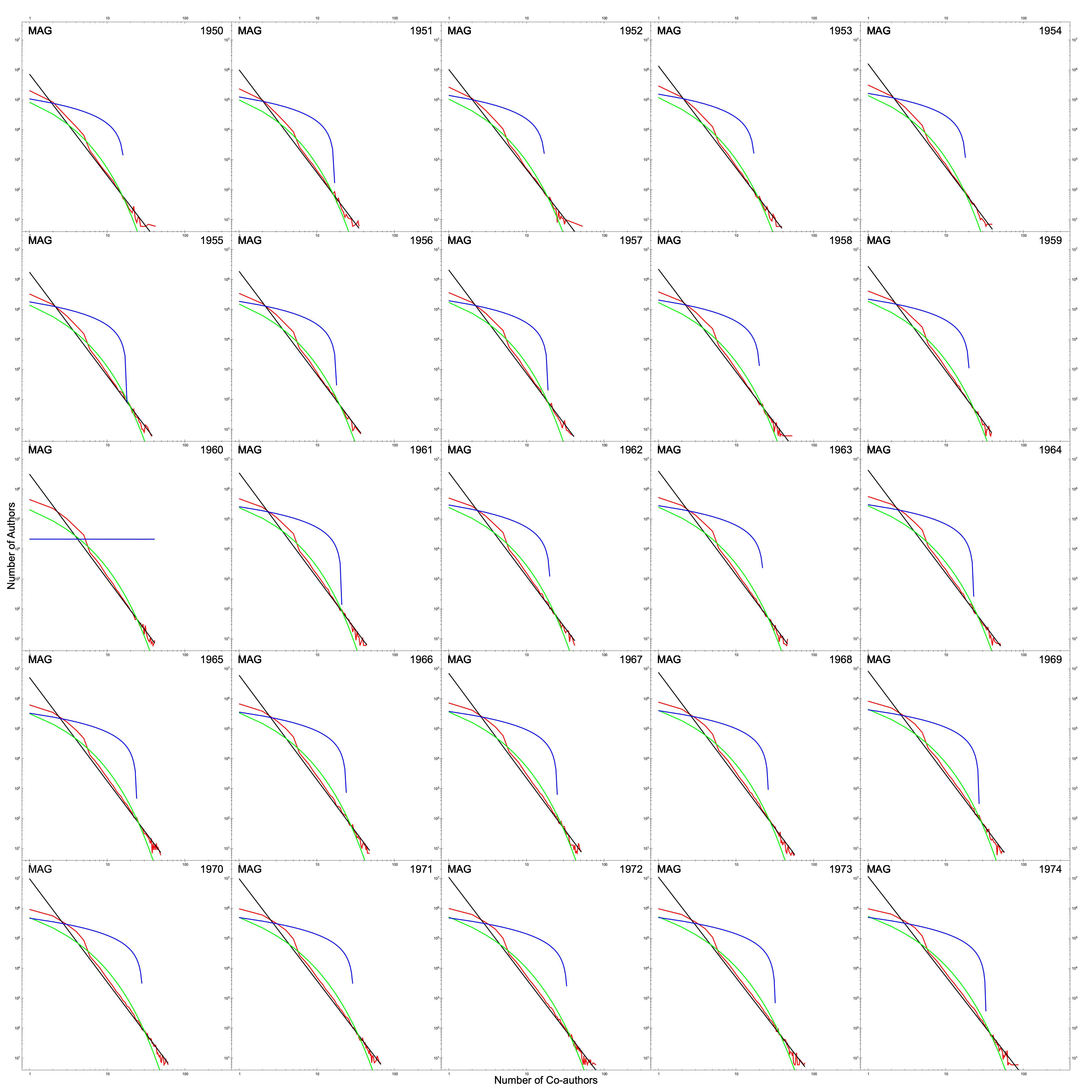"}
  \end{tabular}
  \caption{Degree distributions for cohorts of authors who first published in a given year, 1950 to 1974 (red line). Power-law, Log-normal, and Weibull fits are shown with black, blue, and green lines respectively.}
  \label{mag_degrees_g}
\end{figure*}

\begin{figure*}[pt]
  \centering
  \begin{tabular}{c}
    \includegraphics[width=0.96\textwidth]{"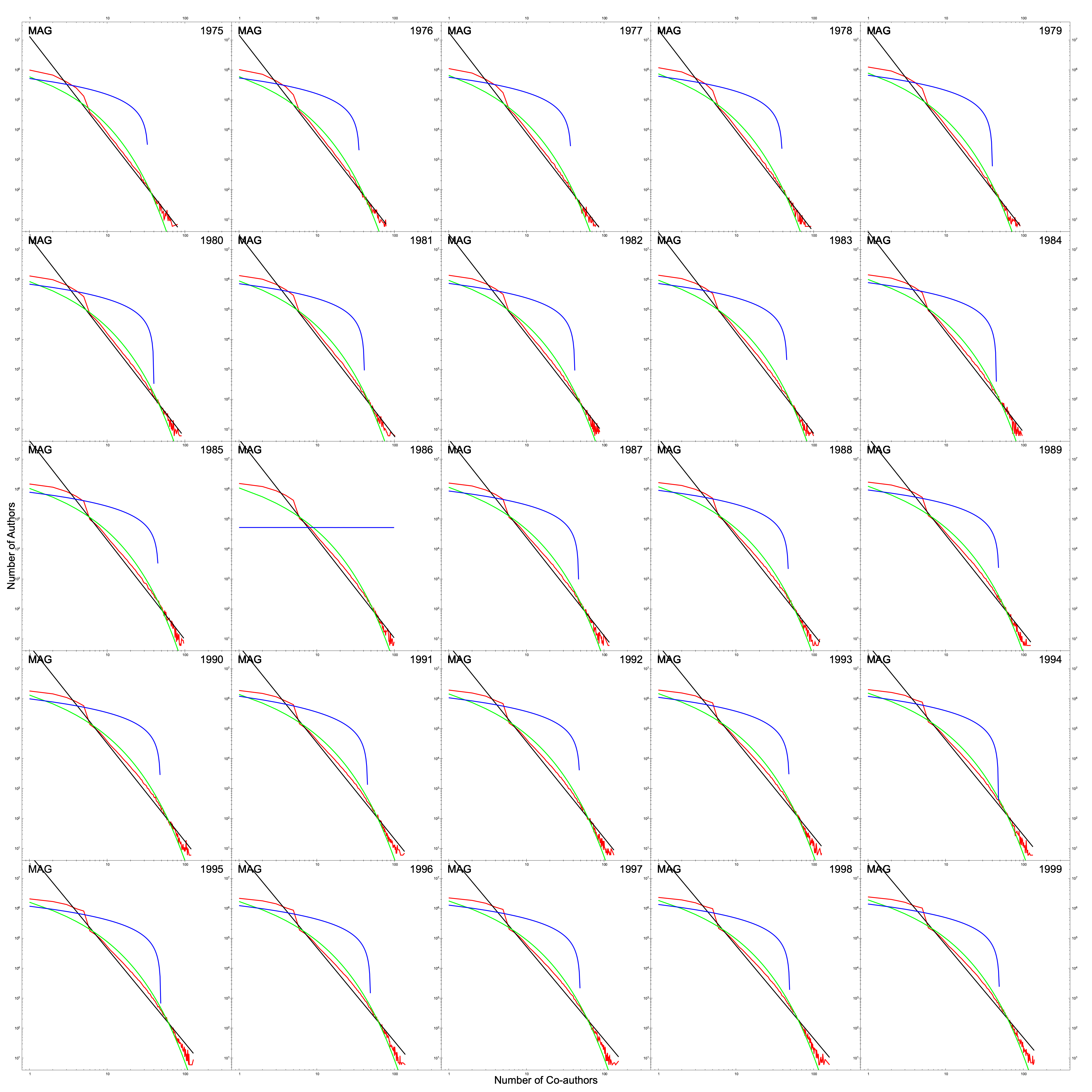"}
  \end{tabular}
  \caption{Degree distributions for cohorts of authors who first published in a given year, 1975 to 1999 (red line). Power-law, Log-normal, and Weibull fits are shown with black, blue, and green lines respectively.}
  \label{mag_degrees_h}
\end{figure*}

\begin{figure*}[pt]
  \centering
  \begin{tabular}{c}
    \includegraphics[width=0.96\textwidth]{"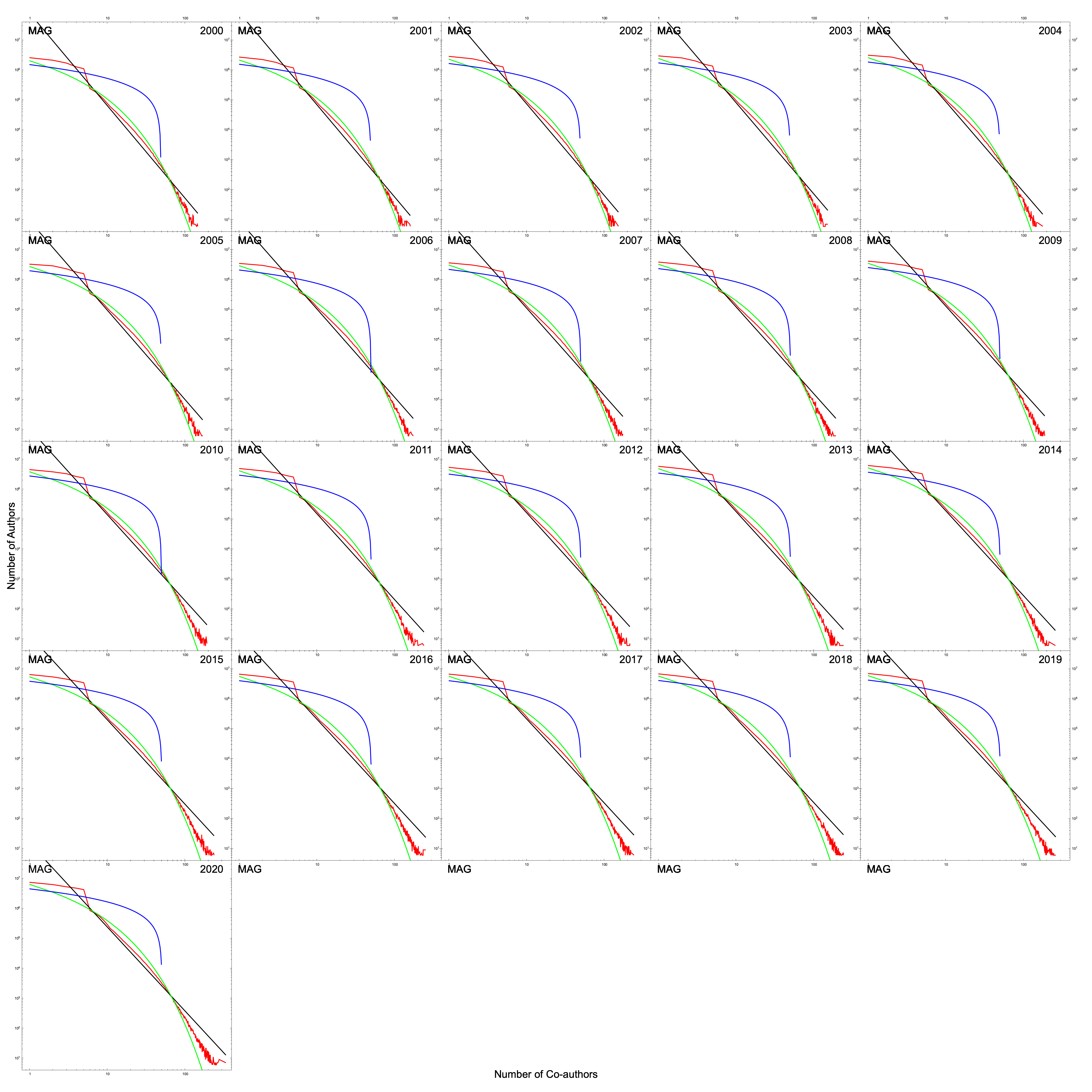"}
  \end{tabular}
  \caption{Degree distributions for cohorts of authors who first published in a given year, 2000 to 2020 (red line). Power-law, Log-normal, and Weibull fits are shown with black, blue, and green lines respectively.}
  \label{mag_degrees_i}
\end{figure*}

\begin{figure*}[pt]
  \centering
  \begin{tabular}{c}
    \includegraphics[width=0.96\textwidth]{"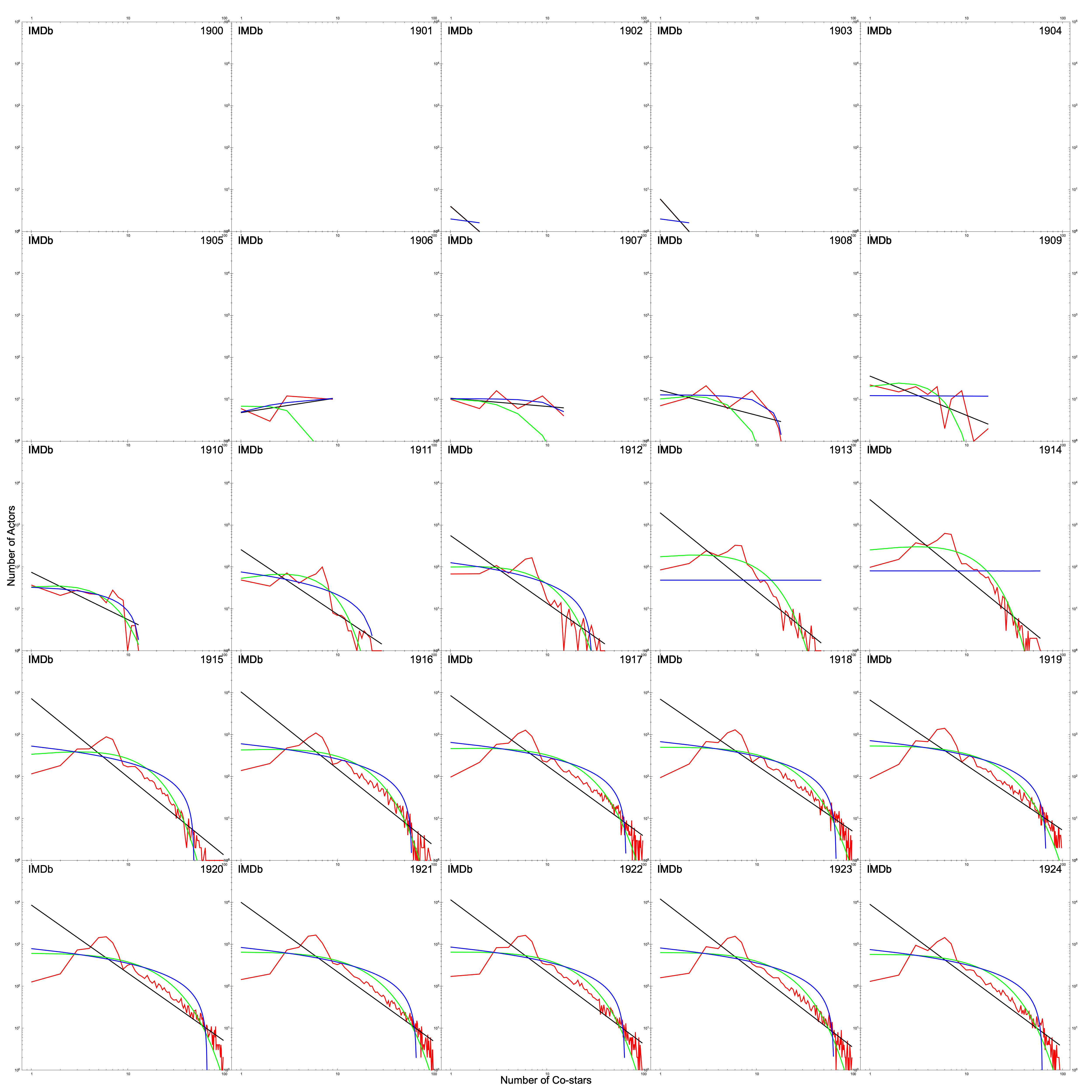"}
  \end{tabular}
  \caption{Degree distributions for cohorts of lead actors who made their first movie in a given year, 1900 to 1924 (red line). Power-law, Log-normal, and Weibull fits are shown with black, blue, and green lines respectively.}
  \label{imdb_degrees_a}
\end{figure*}

\begin{figure*}[pt]
  \centering
  \begin{tabular}{c}
    \includegraphics[width=0.96\textwidth]{"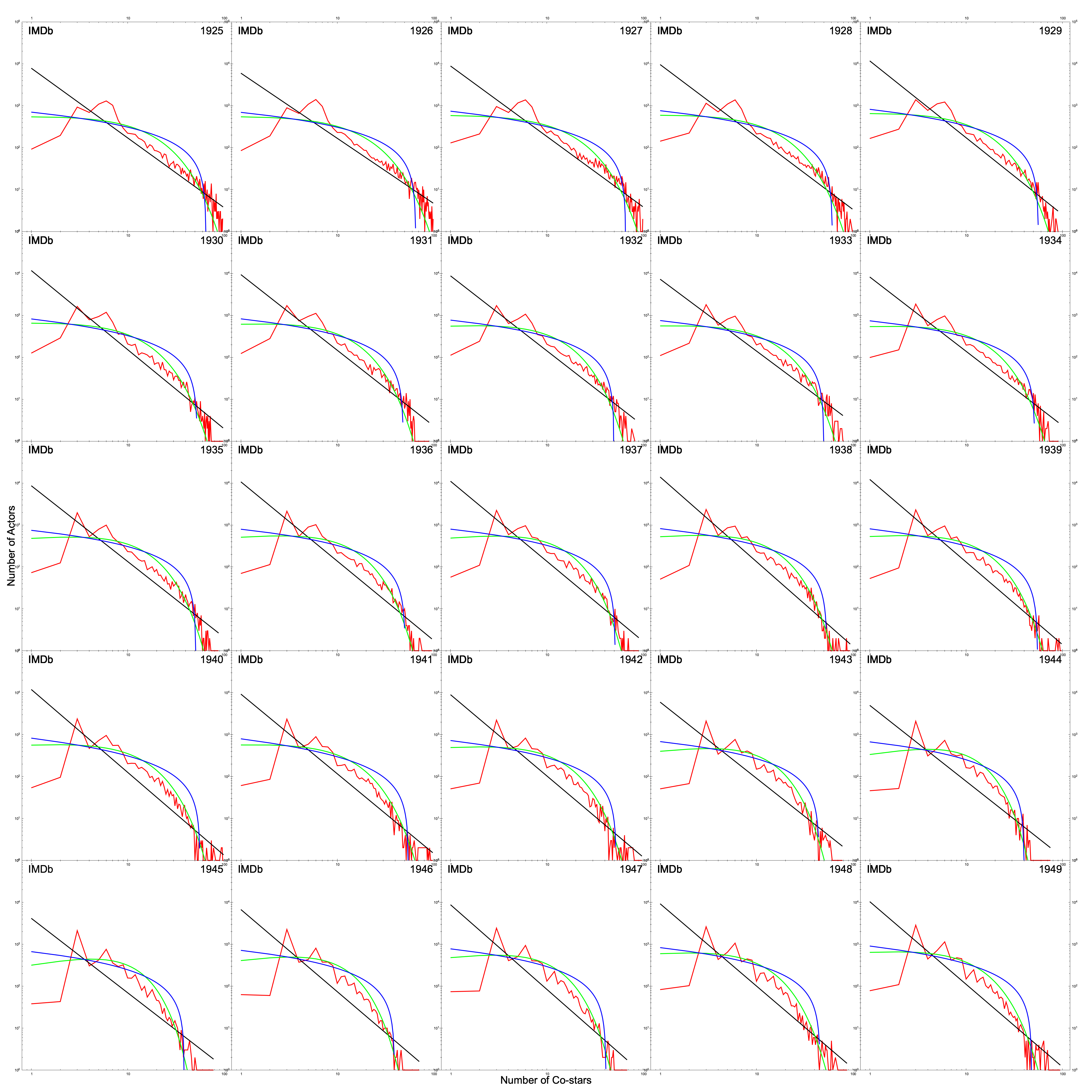"}
  \end{tabular}
  \caption{Degree distributions for cohorts of lead actors who made their first movie in a given year, 1925 to 1949 (red line). Power-law, Log-normal, and Weibull fits are shown with black, blue, and green lines respectively.}
  \label{imdb_degrees_b}
\end{figure*}

\begin{figure*}[pt]
  \centering
  \begin{tabular}{c}
    \includegraphics[width=0.96\textwidth]{"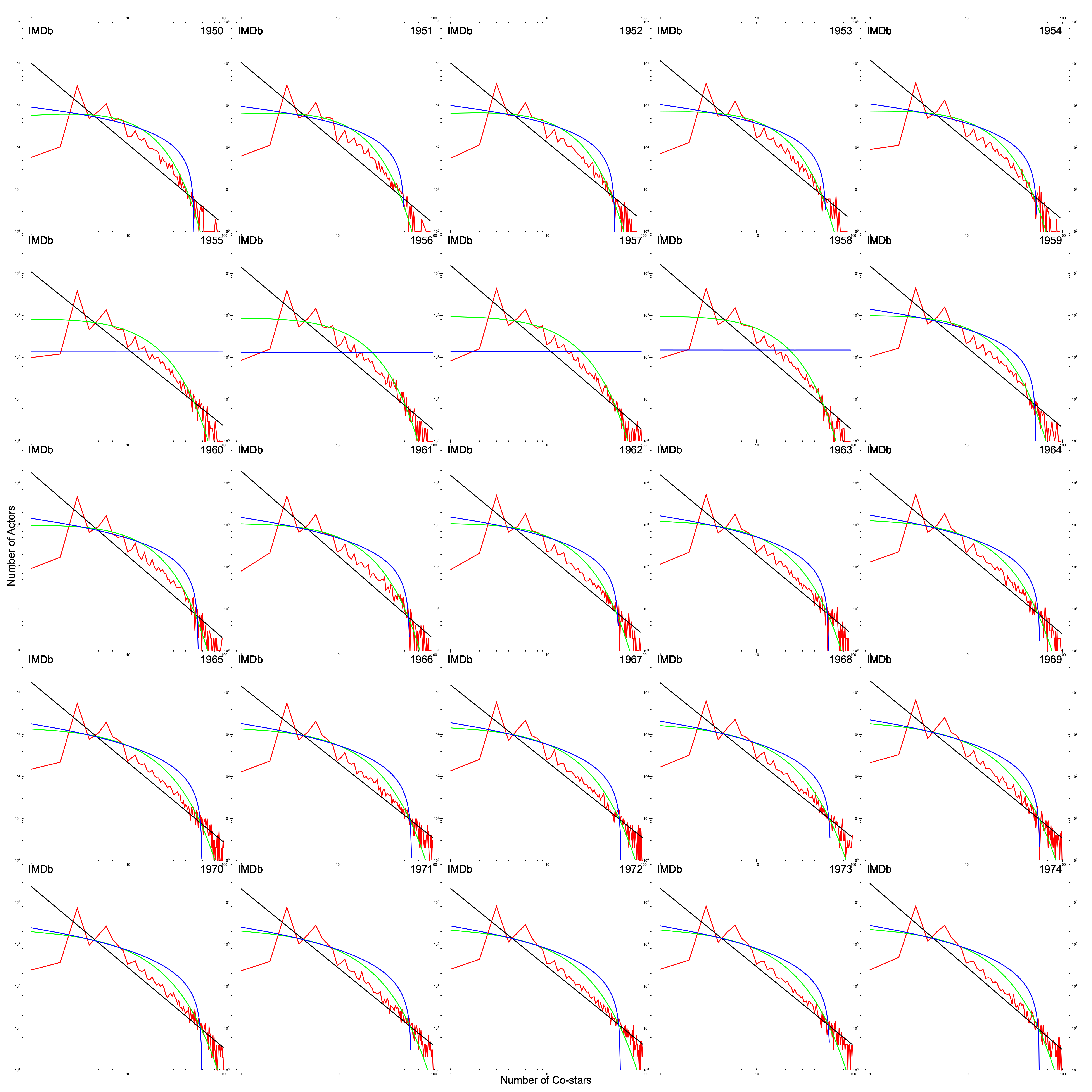"}
  \end{tabular}
  \caption{Degree distributions for cohorts of lead actors who made their first movie in a given year, 1950 to 1974 (red line). Power-law, Log-normal, and Weibull fits are shown with black, blue, and green lines respectively.}
  \label{imdb_degrees_c}
\end{figure*}

\begin{figure*}[pt]
  \centering
  \begin{tabular}{c}
    \includegraphics[width=0.96\textwidth]{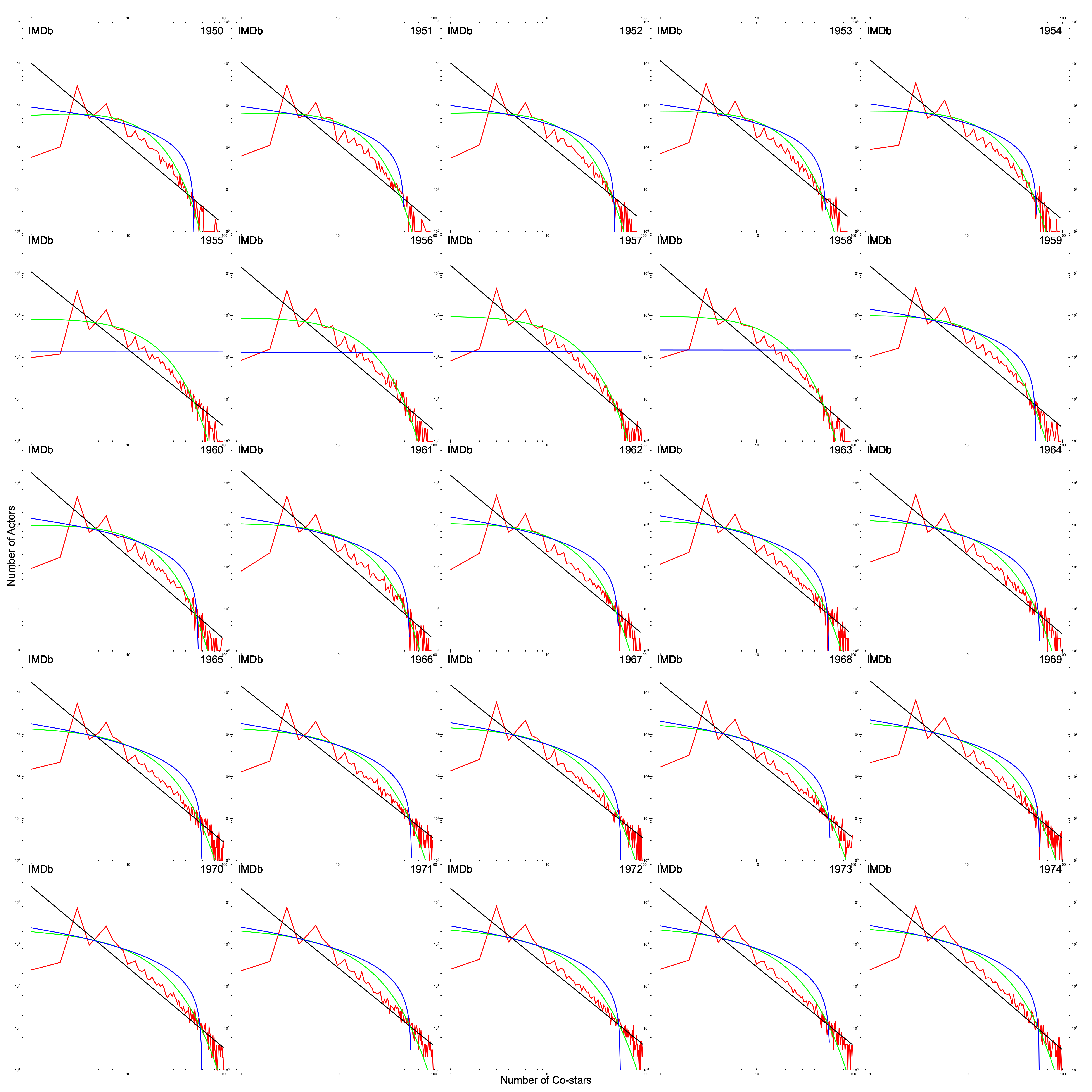}
  \end{tabular}
  \caption{Degree distributions for cohorts of lead actors who made their first movie in a given year, 1975 to 1999 (red line). Power-law, Log-normal, and Weibull fits are shown with black, blue, and green lines respectively.}
  \label{imdb_degrees_d}
\end{figure*}

\begin{figure*}[pt]
  \centering
  \begin{tabular}{c}
    \includegraphics[width=0.96\textwidth]{"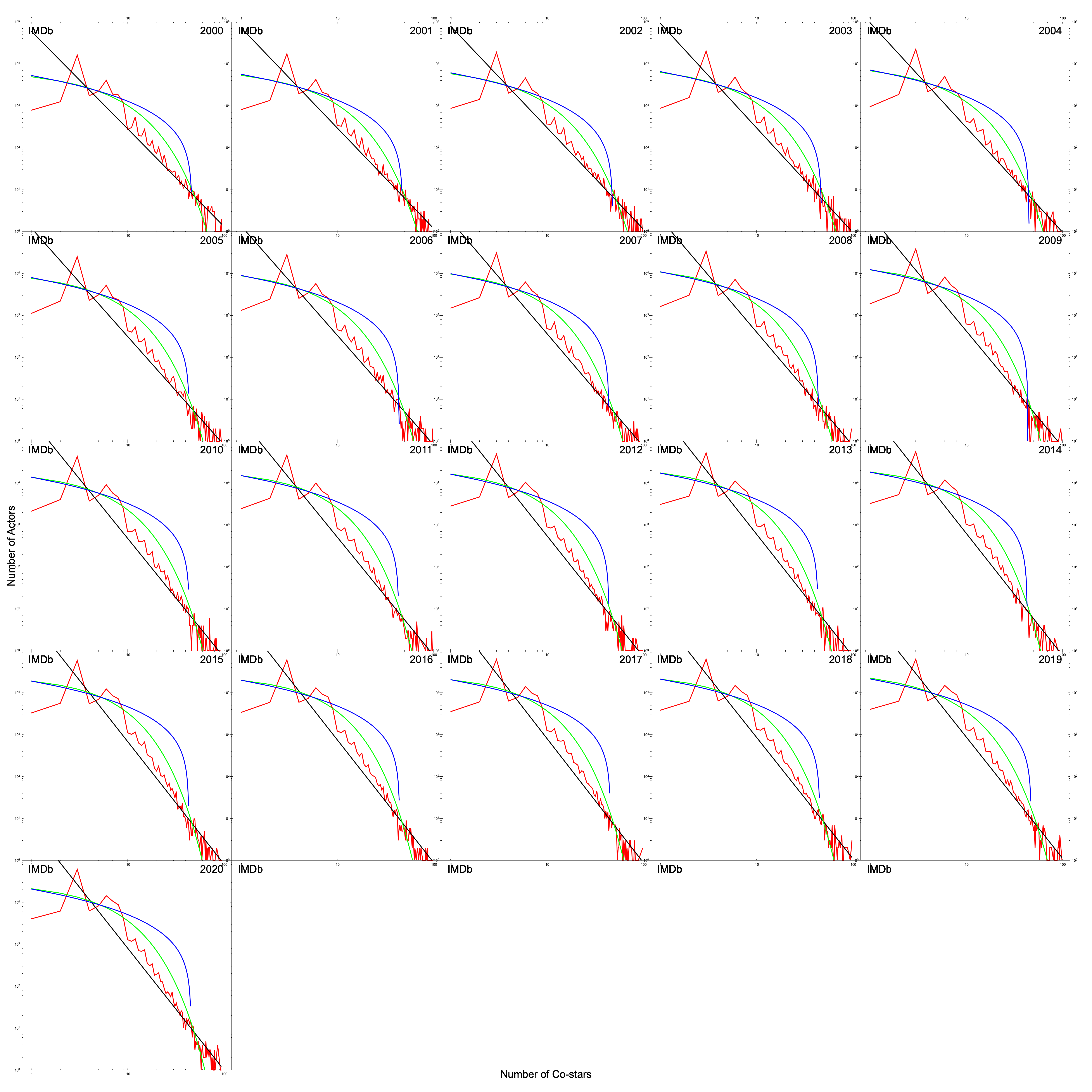"}
  \end{tabular}  
  \caption{Degree distributions for cohorts of lead actors who made their first movie in a given year, 2000 to 2020 (red line). Power-law, Log-normal, and Weibull fits are shown with black, blue, and green lines respectively.}
  \label{imdb_degrees_e}
\end{figure*}

The degree distributions of both networks exhibit distinct evolutionary patterns over their respective timescales: see Figures \ref{mag_degrees_a} to \ref{imdb_degrees_e}. In the MAG network (1800-2020), early distributions (1800-1850) closely follow power-law behaviour, with deviations from power-law fits falling within 5\% across the full degree range. As the network evolves, a characteristic flattening emerges in the low-degree region, quantified by systematic positive residuals from power-law fits reaching 25-35\% for degrees 1-5 by 1900. We define this flattening quantitatively as the percentage excess of observed frequencies over power-law predictions in the low-degree region.

The transition from power-law to Weibull behaviour occurs gradually between 1850-1900, coinciding with network growth from $10^4$ to $10^6$ nodes. During this transition period, the ratio of observed-to-predicted frequencies in the low-degree region ($d \leq 5$) increases monotonically at a rate of approximately $0.5\%$ per year. The cutoff point $d_c$ between the flattened region and power-law tail, defined as the degree at which observed frequencies return to within $5\%$ of power-law predictions, shifts from $d_c \approx 5$ (1850) to $d_c \approx 9$ (1900). The high-degree tail maintains approximate power-law scaling throughout, with deviations $< 10\%$ for degrees $d > d_c$.

The IMDb network (1900-2020) displays a distinctive degree distribution structure that emerges rapidly during its early evolution (1900-1920). The distribution is characterised by a power-law tail ($d > 10$), a pronounced peak at degree $d = 2$ ($40\%$ above power-law predictions), and systematically lower values at degree $d = 1$ ($30\%$ below predictions). This structure reflects domain-specific constraints, as movies typically require 2-3 lead actors for narrative purposes. These features stabilise by 1920 and remain constant ($\pm 5\%$ variation) throughout subsequent evolution.

\subsection{Network Size and Distribution Stability}

To investigate the role of network size in distribution stability, we analysed how fitting parameters vary with node count $N$. For the MAG network, we observe three distinct scaling regimes:
\begin{enumerate}
\item Early Growth ($N < 10^4$): Power-law fits dominate ($\chi^2$ values $10$-$15\%$ lower than Weibull), with unstable parameters (standard deviations $> 20\%$),
\item Transition ($10^4 < N < 10^6$): Gradual shift to Weibull dominance, with decreasing parameter variability,
\item Mature Phase ($N > 10^6$): Stable Weibull parameters despite continued growth.
\end{enumerate}
The Weibull shape parameter $k$ shows strong size dependence during early growth ($k \propto N^{0.15}$ for $N < 10^4$), but stabilises in the range $0.8$-$1.0$ ($\pm 0.05$) once $N$ exceeds $10^6$, remaining constant despite further growth to $10^8$ nodes. The scale parameter $\lambda$ continues to grow as $\lambda \propto \ln(N)$, reflecting the network's expanding degree range.

The IMDb network, despite its smaller size range ($10^3$ to $10^6$ nodes), shows similar size-dependent behaviour. Weibull parameters stabilise at $N \approx 10^5$, with $k = 0.9$-$1.1$ ($\pm 0.07$) thereafter. This suggests a critical network size for distribution stability that may be domain-dependent.

\begin{figure*}[!t]
  \centering
  \begin{tabular}{cc}
    \includegraphics[width=0.47\textwidth]{"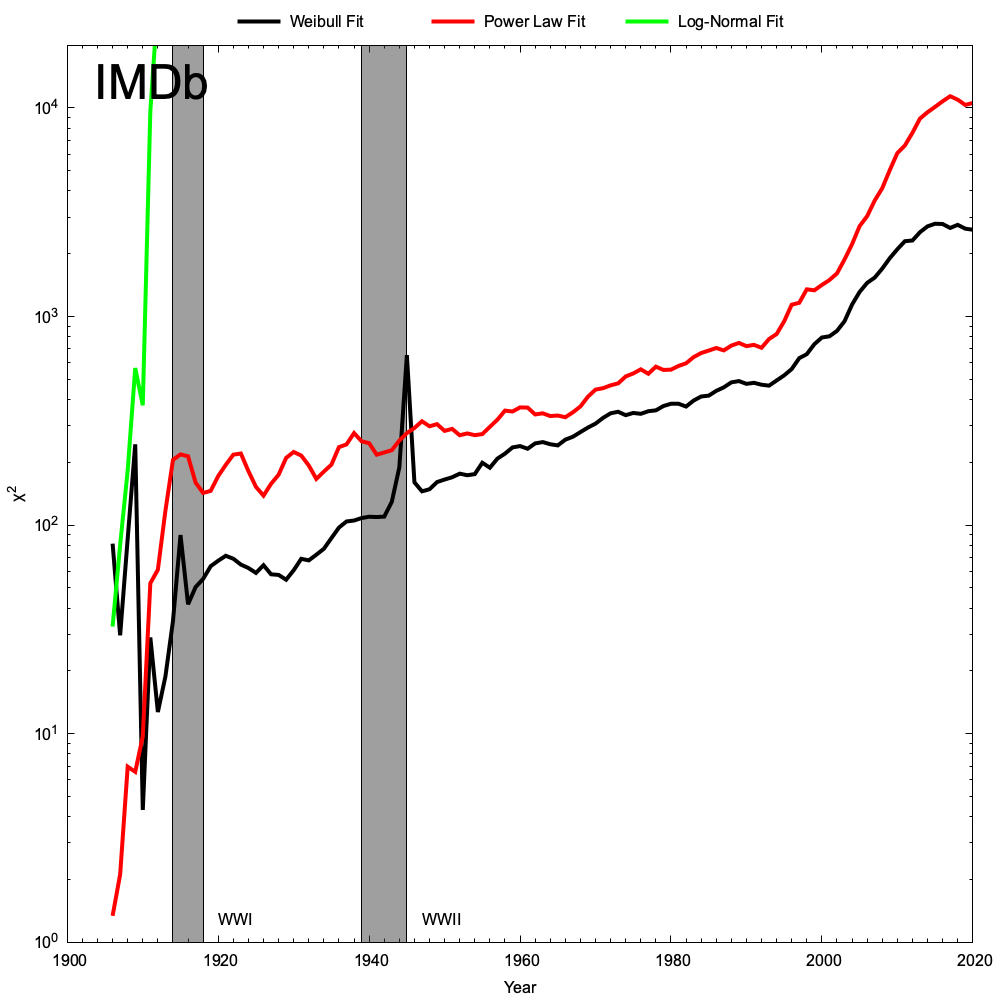"} &
    \includegraphics[width=0.47\textwidth]{"degree-distribution-chi2.png"}
  \end{tabular}
  \caption{$\chi^{2}$ for power-law, log-normal, and Weibull fits to the MAG and IMDb network degree distributions in red, green, and black lines respectively.}
  \label{chi2}
\end{figure*}

\subsection{Statistical Analysis of Distribution Fits}

Figures and \ref{degree_fits} and \ref{chi2} show the evolution of the fitting parameters, allowing us to perform a statistical comparison of power-law, log-normal, and Weibull distributions across all time periods. In the MAG network, power-law fits show marginally better performance during early periods (1800-1850), with $\chi^2$ values 10-15\% lower than Weibull fits. After 1850, Weibull distributions consistently provide superior fits, with $\chi^2$ values 20-30\% lower than power-law fits and 40-50\% lower than log-normal fits.

The IMDb network presents even stronger evidence for Weibull behaviour. From its earliest periods, Weibull distributions outperform both alternatives, with $\chi^2$ values typically 30-40\% lower than power-law fits. Log-normal distributions consistently provide the poorest fits across all periods in both networks. This suggests that pure multiplicative growth processes do not adequately describe collaboration network evolution.

\subsection{Stability of Distribution Parameters}

Both networks show remarkable stability in their Weibull shape parameters despite dramatic growth in network size. In the MAG network, the Weibull shape parameter $k$ remains consistently in the range 0.8-1.0 from 1850 onwards, even as the network grows from approximately $10^4$ to $10^8$ nodes. The scale parameter $\lambda$ increases systematically with network growth, reflecting the expansion in typical degree values.

The IMDb network exhibits similar parameter stability, with $k$ values concentrated in the range 0.9-1.1 throughout its evolution. The two distinct regions of the MAG degree distribution become increasingly well-defined in the post-WWII period. This coincides with the significant increase in multi-authorship patterns previously reported \citep{williams2025}.

\subsection{Temporal Evolution}

The degree distribution parameters show a systematic temporal evolution. The MAG network's power-law index $\gamma$ decreases from approximately 2.8 to 2.3 between 1800 and 1900, then stabilises. The Weibull shape parameter $k$ reaches its characteristic range by 1850 and maintains this range through 2020. The Weibull scale parameter $\lambda$ exhibits distinct growth phases, with accelerated increases during periods of network expansion.

Despite their different domains, both networks maintain stable functional forms characterised by Weibull distributions. Neither power-law nor log-normal distributions adequately capture the complete degree distributions, particularly in the low-degree region. The consistency of Weibull parameters across these distinct collaboration domains, spanning different time periods and scales of growth, suggests that Weibull distributions may capture universal aspects of collaboration network formation.

This stability in distribution form persists despite the MAG network's significant shift toward multi-authorship in the post-WWII period. These findings imply that while collaboration patterns may change, the underlying processes generating degree distributions maintain consistent statistical properties. The improved statistical agreement with Weibull distributions suggests that collaboration networks may be shaped more by systemic constraints on partnership formation than by unconstrained preferential attachment or multiplicative growth processes.

\section{Discussion}

This analysis of degree distributions in large collaboration networks has revealed patterns that extend beyond the classic preferential attachment framework. While \citet{barabasi1999} demonstrated how power-law distributions can emerge from preferential attachment, these collaboration networks appear to follow more intricate growth patterns. The $\chi^2$ goodness-of-fit measures favoured Weibull distributions, and suggests that the formation of collaborative ties may be governed by additional factors beyond simple rich-get-richer dynamics.

While the chi-squared goodness-of-fit test provides a robust and widely used measure for comparing distributions, it is important to acknowledge its limitations. As a binning-based method, the resulting statistic can be influenced by bin choices, and it might not capture subtle differences in distribution shape as effectively as more nuanced statistical tests. Indeed, a variety of alternative goodness-of-fit measures exist, such as the Kolmogorov-Smirnov statistic, likelihood ratio tests, and information criteria like AIC and BIC, which offer different sensitivities and strengths. However, given the scale and temporal resolution of our network data, employing highly sensitive or complex statistical measures may be inappropriate, potentially leading to overfitting to noise or minor fluctuations. Moreover, with 1000 nodes per bin, the chi-squared statistic benefits from consistent statistical power across the entire distribution range. In this context, the chi-squared test offers a pragmatic balance, providing a sufficiently robust and interpretable metric to reveal the main trends in distributional evolution and to compare the general fit of power-law, log-normal, and Weibull models, which is the primary focus of our analysis.

The consistent Weibull shape parameters ($k \approx 0.8$-$1.0$ for MAG, $k \approx 0.9$-$1.1$ for IMDb) suggest a possible universal mechanism underlying collaboration network formation. Similar Weibull-like behaviour appears in firm size distributions \citep{cabral2003}, scientific citation patterns \citep{golosovsky2017}, and income distributions \citep{ghosh2014}. Recent work by \citet{zhou2018} found comparable patterns in online social networks, suggesting Weibull processes may be fundamental to social organisation.

These results can be viewed in the context of existing theoretical frameworks for network growth, which suggest possible mechanisms for the observed patterns. Also, the consistent Weibull distributions observed in both collaboration networks suggest an underlying dynamical process that can be understood through statistical physics. In physical systems, Weibull distributions typically emerge from processes with competing exponential rates \citep{sornette2006}. The observed network evolution appears to reflect similar competing processes: exponential growth through preferential attachment, exponential decay in connection probability due to resource constraints, and power-law scaling from hierarchical organization. This interplay can be formalized through a generalized growth equation:
\begin{equation}
\frac{dk}{dt} = \alpha k^\beta e^{-\gamma k},
\label{eq:growth}
\end{equation}
where $k$ is node degree, and ${\alpha, \beta, \gamma}$ are parameters reflecting growth rate, preferential attachment strength, and resource constraints, respectively.

This growth equation belongs to a broader class of competitive dynamical systems studied in population dynamics \citep{may1976}, where stable distributions emerge from the balance between growth and limiting factors. Such systems typically approach stable fixed points in the presence of multiplicative noise \citep{gardiner2009}, generating Weibull-like distributions similar to those we observe in both the academic and entertainment networks. Recent theoretical work strengthens this interpretation: \citet{krioukov2016} showed how geometric constraints naturally generate non-power-law degree distributions, while \citet{bianconi2001} demonstrated how resource competition modifies preferential attachment behavior. The stable Weibull parameters we observe ($k \approx 0.8$-$1.0$ for academic, $k \approx 0.9$-$1.1$ for entertainment collaborations) suggest these networks operate in a regime where growth-limiting constraints dominate pure preferential attachment processes.

The identification of Weibull processes in collaboration networks has practical implications. In organisational design, network architecture should account for natural degree constraints and cognitive limitations in team formation. Research policy can benefit from understanding natural collaboration size limits and domain-specific degree constraints when designing funding mechanisms and evaluation metrics. Social and professional networking platforms could optimise their design by accommodating these natural constraints rather than encouraging unlimited connection growth. Recent work by \citet{seshadri2021} shows how understanding degree distribution shape can improve network resilience prediction and resource allocation efficiency.

The broader implications of our findings contribute to fundamental debates about universality in complex systems. The emergence of Weibull distributions in collaboration networks suggests that constrained growth processes may be as universal as the critical phenomena that generate power laws. This observation challenges the common assumption that social networks naturally evolve toward increasingly centralised, scale-free structures. Instead, our results suggest that robust constraint mechanisms naturally limit such concentration, potentially contributing to network stability and resilience.

The relationship between microscopic interaction rules and macroscopic Weibull behaviour requires deeper theoretical investigation. While we have proposed a general framework based on competing exponential processes, the precise mechanisms translating individual collaborative decisions into network-level degree distributions remain unclear. \citet{voitalov2019} provides mathematical tools for more rigorous distribution analysis, but bridging the gap between individual behaviour and network statistics presents significant challenges. The framework of \citet{wang2021} suggests that growth constraints might emerge from optimisation principles, offering one potential theoretical direction.

Integration with existing network theory raises intriguing questions. The stability of Weibull parameters despite explosive network growth suggests a form of self-organisation distinct from the critical phenomena typically associated with power laws. This observation connects to recent work questioning the universality of scale-free networks \citep{broido2019} and may suggest a more general class of constraint-driven network evolution processes. The temporal stability we observe also relates to fundamental questions about network homeostasis and regulation \citep{holme2012}.

A promising direction for future work involves the relationship between node-level constraints and network-level structure. Recent developments in higher-order network analysis \citep{benson2018} provide tools for investigating how local collaboration constraints propagate through the network hierarchy. These methods reveal how microscopic constraints can generate consistent macroscopic patterns across multiple scales of organisation. The role of institutional structures in mediating between individual and network-level behaviour appears important in maintaining stable Weibull-like distributions, despite significant environmental changes.

Emerging technologies offer new opportunities to test and refine our understanding. High-resolution digital collaboration platforms could provide data on collaboration formation at previously inaccessible temporal and structural scales. The increasing availability of metadata about collaboration context and content could help identify specific constraints shaping network evolution. Online platforms enable experimental manipulation of collaboration constraints, potentially allowing direct testing of causal mechanisms. Such experiments could help determine whether Weibull distributions emerge naturally from human cognitive and social limitations or reflect technological and institutional constraints.

The implications of these findings extend beyond network science to questions in complex systems theory. The emergence of consistent statistical patterns across both academic and entertainment collaboration networks hints at common principles underlying social organization, though conclusions about universality cannot be drawn from just two cases. Previous work suggests these patterns may emerge from the interplay between human cognitive capacity \citep{dunbar2016}, institutional structures \citep{newman2001b}, and resource limitations \citep{guimera2005}. Understanding how these constraints shape network evolution could inform theories of human social organization across multiple scales, from small teams to global collaboration networks.

Our findings also provide practical guidance for network intervention strategies. Traditional approaches often aim to maximise connectivity or centralise information flow. However, our results suggest successful interventions should instead focus on maintaining natural constraints while optimising within them. This perspective aligns with recent work on sustainable network design and could inform policies aimed at promoting healthy collaboration patterns in both scientific and entertainment communities.

The journey from empirical observation of Weibull distributions to fundamental understanding of collaboration networks remains incomplete. However, our findings establish a foundation for future investigation while raising important questions about the nature of human collaborative systems. As network science continues to mature, the role of node-level and edge-level constraints in shaping the macroscopic network evolution may prove as fundamental as the growth processes that have dominated theoretical attention to date.

\section{Conclusions}

In this paper we have analysed two large-scale collaboration networks, each extending over a century. The central finding is that Weibull distributions outperform power-law and log-normal fits across many decades of data in both the MAG and IMDb networks. This finding contributes to a growing body of evidence \citep{broido2019, voitalov2019, seshadri2021} suggesting that real-world networks often deviate systematically from pure power-law behaviour, particularly in systems constrained by human cognitive and social limitations. These constraints appear to fundamentally shape network evolution through limitations on collaboration size and connectivity: the development of a precise generative model to explain their emergence and effects remains an important direction for future work.

The Microsoft Academic Graph (1800-2020) and Internet Movie Database (1900-2020) networks show remarkably consistent Weibull behaviour, with shape parameters stabilising at $k \approx 0.8$-$1.0$ and $k \approx 0.9$-$1.1$ respectively. This stability persists despite dramatic growth in network size (from $10^4$ to $10^8$ nodes in MAG) and significant changes in collaboration patterns. The consistency across such different domains suggests universal constraints on collaborative behaviour that transcend specific institutional contexts.

Both networks exhibit distinct evolutionary patterns that illuminate the transition from early growth to mature structure. Early-stage networks show approximate power-law scaling, particularly evident in the MAG network before 1850. As the networks mature, they develop characteristic flattening in the low-degree region that Weibull distributions capture accurately. The cutoff between the flattened region and power-law tail systematically increases from approximately 5 to 9 edges in the MAG network, while the IMDb network maintains a distinctive degree structure with a stable peak at degree 2, reflecting domain-specific narrative constraints.

Statistical analysis revealed that Weibull distributions provide better fits across most time periods. In mature networks, Weibull fits outperform power-law distributions by 20-30\% in the MAG network and 30-40\% in the IMDb network according to $\chi^2$ metrics. This improved fit persists under multiple statistical tests and remains robust to significant perturbations in network structure, including the post-WWII shift toward increased multi-authorship in academic collaboration.

These results suggest that collaboration network evolution may be governed by constraint-based growth, rather than pure preferential attachment or multiplicative processes. The stability of Weibull parameters across domains suggests these constraints may reflect fundamental limitations on human collaborative capacity. However, we emphasise that distribution fitting alone cannot definitively identify generating mechanisms. Similar distribution patterns may arise from various distinct processes, highlighting the importance of direct measurement of growth processes and theoretical modelling to better understand these relationships.

Our findings have important implications beyond network science, suggesting universal principles in how human collaboration systems organise and evolve. The emergence of stable Weibull distributions suggests that sustainable networks naturally balance growth processes against cognitive and resource constraints. This understanding could inform practical applications from organisational design to research policy, particularly in creating environments that work with rather than against natural collaborative tendencies.

Further theoretical work is needed to understand how local constraints generate global Weibull behaviour, particularly through the lens of competitive dynamical systems. Analysis of other social and technological networks could test the generality of our findings, while higher-resolution temporal data might reveal a finer structure in collaboration patterns. Finally, investigation of the relationship between node-level constraints and network-level structure could advance our understanding of complex system organisation across multiple scales.

The robustness of Weibull behaviour in these two large collaboration networks suggests we may be approaching a more complete understanding of how human social systems naturally organise and evolve. This insight offers a foundation for designing more effective and sustainable collaborative systems, aligned with rather than opposed to the fundamental constraints that shape human social organisation.

\begin{acks}
This research was carried out at Rinna K.K., Tokyo, Japan.
\end{acks}
\bibliographystyle{SageH}
\bibliography{paper3}

\begin{thebibliography}{34}
\providecommand{\natexlab}[1]{#1}
\providecommand{\url}[1]{\texttt{#1}}
\providecommand{\urlprefix}{URL }
\expandafter\ifx\csname urlstyle\endcsname\relax
  \providecommand{\doi}[1]{DOI:\discretionary{}{}{}#1}\else
  \providecommand{\doi}{DOI:\discretionary{}{}{}\begingroup
  \urlstyle{rm}\Url}\fi

\bibitem[{Albert et~al.(1999)Albert, Jeong and Barab\'{a}si}]{albert1999}
Albert R, Jeong H and Barab\'{a}si AL (1999) Diameter of the {World-Wide Web}.
\newblock \emph{Nature} 401: 130--131.
\newblock \doi{10.1038/43601}.

\bibitem[{Barab\'{a}si and Albert(1999)}]{barabasi1999}
Barab\'{a}si AL and Albert R (1999) Emergence of scaling in random networks.
\newblock \emph{Science} 286: 509--512.

\bibitem[{Benson et~al.(2018)Benson, Gleich and Leskovec}]{benson2018}
Benson AR, Gleich DF and Leskovec J (2018) Higher-order organization of complex
  networks.
\newblock \emph{Science} 353: 163--166.

\bibitem[{Bianconi and Barab\'{a}si(2001)}]{bianconi2001}
Bianconi G and Barab\'{a}si AL (2001) Competition and multiscaling in evolving
  networks.
\newblock \emph{Europhysics Letters} 54: 436--442.

\bibitem[{Broido and Clauset(2019)}]{broido2019}
Broido AD and Clauset A (2019) Scale-free networks are rare.
\newblock \emph{Nature Communications} 10: 1017.

\bibitem[{Cabral and Mata(2003)}]{cabral2003}
Cabral LMB and Mata J (2003) On the evolution of the firm size distribution:
  Facts and theory.
\newblock \emph{American Economic Review} 93: 1075--1090.

\bibitem[{Clauset et~al.(2009)Clauset, Shalizi and Newman}]{clauset2009}
Clauset A, Shalizi CR and Newman MEJ (2009) Power-law distributions in
  empirical data.
\newblock \emph{SIAM Review} 51: 661--703.

\bibitem[{Dunbar(2016)}]{dunbar2016}
Dunbar RIM (2016) Do online social media cut through the constraints that limit
  the size of offline social networks?
\newblock \emph{Royal Society Open Science} 3: 150292.

\bibitem[{Gardiner(2009)}]{gardiner2009}
Gardiner C (2009) \emph{Handbook of Stochastic Methods}.
\newblock 4th edition. Berlin: Springer.

\bibitem[{Ghosh et~al.(2014)Ghosh, Chattopadhyay and Chakrabarti}]{ghosh2014}
Ghosh A, Chattopadhyay N and Chakrabarti BK (2014) Inequality in societies,
  academic institutions and science journals: Gini and k-indices.
\newblock \emph{Physica A} 410: 30--34.

\bibitem[{Golosovsky and Solomon(2017)}]{golosovsky2017}
Golosovsky M and Solomon S (2017) Growing complex network of citations of
  scientific papers: Modeling and measurements.
\newblock \emph{Physical Review E} 95: 012324.

\bibitem[{Guimer\`{a} et~al.(2005)Guimer\`{a}, Uzzi, Spiro and
  Amaral}]{guimera2005}
Guimer\`{a} R, Uzzi B, Spiro J and Amaral LAN (2005) Team assembly mechanisms
  determine collaboration network structure and team performance.
\newblock \emph{Science} 308: 697--702.

\bibitem[{Holme(2019)}]{holme2019}
Holme P (2019) Rare and everywhere: Perspectives on scale-free networks.
\newblock \emph{Nature Communications} 10: 1016.

\bibitem[{Holme and Saram\"{a}ki(2012)}]{holme2012}
Holme P and Saram\"{a}ki J (2012) Temporal networks.
\newblock \emph{Physics Reports} 519: 97--125.

\bibitem[{IMDb(2023)}]{imdbweb}
IMDb (2023) Imdb datasets.
\newblock \url{https://www.imdb.com/interfaces/}.
\newblock Data downloaded in 2023.

\bibitem[{Jeong et~al.(2001)Jeong, Mason, Barab\'{a}si and Oltvai}]{jeong2001}
Jeong H, Mason SP, Barab\'{a}si AL and Oltvai ZN (2001) Lethality and
  centrality in protein networks.
\newblock \emph{Nature} 411: 41--42.

\bibitem[{Krioukov et~al.(2016)Krioukov, Kitsak, Sinkovits, Rideout, Meyer and
  Bogu{\~n}\'{a}}]{krioukov2016}
Krioukov D, Kitsak M, Sinkovits RS, Rideout D, Meyer D and Bogu{\~n}\'{a} M
  (2016) Network cosmology.
\newblock \emph{Scientific Reports} 6: 33416.

\bibitem[{May(1976)}]{may1976}
May R (1976) Simple mathematical models with very complicated dynamics.
\newblock \emph{Nature} 261: 459--467.

\bibitem[{Microsoft(2015)}]{magweb}
Microsoft (2015) Microsoft academic graph.
\newblock
  \url{https://www.microsoft.com/en-us/research/project/microsoft-academic-graph/}.

\bibitem[{Mitzenmacher(2004)}]{mitzenmacher2004}
Mitzenmacher M (2004) A brief history of generative models for power law and
  lognormal distributions.
\newblock \emph{Internet Mathematics} 1: 226--251.

\bibitem[{Nekola and Brown(2003)}]{nekola2003}
Nekola J and Brown J (2003) The wealth of species: ecological communities,
  complex systems and the legacy of frank preston.
\newblock \emph{Ecology Letters} 6: 265--279.

\bibitem[{Newman(2001{\natexlab{a}})}]{newman2001b}
Newman MEJ (2001{\natexlab{a}}) Scientific collaboration networks: I. network
  construction and fundamental results.
\newblock \emph{Physical Review E} 64: 016131.

\bibitem[{Newman(2001{\natexlab{b}})}]{newman2001}
Newman MEJ (2001{\natexlab{b}}) The structure of scientific collaboration
  networks.
\newblock \emph{PNAS} 98: 404--409.

\bibitem[{Newman(2005)}]{newman2005}
Newman MEJ (2005) \emph{Power laws, Pareto distributions and Zipf's law},
  volume~46.

\bibitem[{Seshadri et~al.(2021)Seshadri, Machta, Clauset and
  Larremore}]{seshadri2021}
Seshadri M, Machta S, Clauset A and Larremore DB (2021) The evolution of
  citation graphs.
\newblock \emph{Nature Communications} 12: 1--11.

\bibitem[{Sinha et~al.(2015)Sinha, Shen, Song, Ma, Eide, Hsu and
  Wang}]{sinha2015}
Sinha A, Shen Z, Song Y, Ma H, Eide D, Hsu BJP and Wang K (2015) An overview of
  microsoft academic service (mas) and applications.
\newblock In: \emph{Proceedings of the 24th International Conference on World
  Wide Web}. pp. 243--246.

\bibitem[{Sornette(2006)}]{sornette2006}
Sornette D (2006) \emph{Critical Phenomena in Natural Sciences}.
\newblock 2nd edition. Berlin: Springer.

\bibitem[{Stumpf and Porter(2012)}]{stumpf2012}
Stumpf MPH and Porter MA (2012) Critical truths about power laws.
\newblock \emph{Science} 335: 665--666.

\bibitem[{Thompson et~al.(2022)Thompson, Thompson and
  Caetano-Anoll\'{e}s}]{thompson2022}
Thompson JR, Thompson BT and Caetano-Anoll\'{e}s G (2022) The growth and
  scaling of biological networks.
\newblock \emph{Journal of Theoretical Biology} 532: 110928.

\bibitem[{Voitalov et~al.(2019)Voitalov, van~der Hoorn, van~der Hofstad and
  Krioukov}]{voitalov2019}
Voitalov I, van~der Hoorn P, van~der Hofstad R and Krioukov D (2019) Scale-free
  networks well done.
\newblock \emph{Physical Review Research} 1: 033034.

\bibitem[{Wang et~al.(2021)Wang, Song and Barab\'{a}si}]{wang2021}
Wang D, Song C and Barab\'{a}si AL (2021) Quantifying long-term scientific
  impact.
\newblock \emph{Science} 342: 127--132.

\bibitem[{Weibull(1951)}]{weibull1951}
Weibull W (1951) A statistical distribution function of wide applicability.
\newblock \emph{Journal of Applied Mechanics} 18: 293--297.

\bibitem[{Williams and Chen(2024)}]{williams2025}
Williams PR and Chen Z (2024) Explosive growth in large-scale collaboration
  networks.
\newblock \emph{Collective Intelligance} XX: XX.

\bibitem[{Zhou et~al.(2018)Zhou, Wang, Chen and Zhang}]{zhou2018}
Zhou Y, Wang X, Chen R and Zhang G (2018) Degree distribution predictability of
  online social networks.
\newblock \emph{ACM Transactions on the Web} 12: 1--24.

\end{thebibliography}

\end{document}